\documentclass[aps,prb,reprint,superscriptaddress,floatfix]{revtex4-1}

\usepackage{natbib}
\usepackage{graphicx}
\usepackage[T1]{fontenc}
\usepackage{nicefrac}
\usepackage{booktabs}
\usepackage{siunitx}
\usepackage{dcolumn}
\usepackage[flushleft]{threeparttable}
\usepackage[version=3]{mhchem}
\usepackage{url}
\usepackage{todonotes}
\usepackage{lipsum}

\begin{document}
	
	\title{Ab initio Investigation of Structural Stability and Exfoliation 
	Energies in Transition Metal Dichalcogenides based on Ti-, V-, and Mo-Group 
	Elements}
	
	\author{Carlos M. O. Bastos}
	\affiliation{S\~ao Carlos Institute of Physics, University of S\~ao Paulo, 
	PO Box 369, 13560-970, S\~ao Carlos, SP, Brazil.}
	
        \author{Rafael Besse}
	\affiliation{S\~ao Carlos Institute of Physics, University of S\~ao Paulo, 
	PO Box 369, 13560-970, S\~ao Carlos, SP, Brazil.}
    
	\author{Juarez L. F. Da Silva}
	\affiliation{S\~ao Carlos Institute of Chemistry, University of S\~ao 
	Paulo, PO Box 780, 13560-970, S\~ao Carlos, SP, Brazil.}
	
	\author{Guilherme M. Sipahi}
	\affiliation{S\~ao Carlos Institute of Physics, University of S\~ao Paulo, 
	PO Box 369, 13560-970, S\~ao Carlos, SP, Brazil.}

\begin{abstract}
In this work, we report an \textit{ab initio} investigation based on density functional 
theory of the structural, energetic and electronic properties of 2D layered chalcogenides 
compounds based in the combination of the transition-metals (\ce{Ti}, \ce{Zr}, 
\ce{Hf}, \ce{V}, \ce{Nb}, \ce{Ta}, \ce{Cr}, \ce{Mo}, \ce{W} and chalcogenides (\ce{S}, 
\ce{Se}, \ce{Te}) in three polymorphic phases: trigonal prismatic (2H), octahedral 
(1T) and distorted octahedral (1T$_{\text{d}}$). We determined the most stable 
phases for each compound, verifying the existence of the 1T$_{\text{d}}$ phase 
for a small number of the compounds and we have also identified the magnetic compounds. 
In addition, with the determination of the exfoliation energies, we indicated the 
potential candidates to form one layer material and we have also found a relation 
between the exfoliation energy and the effective Bader charge in the metal, suggesting 
that when the materials present small exfoliation energy, it is due to the Coulomb 
repulsion between the chalcogen planes. Finally, we analyzed the electronic properties, 
identifying the semiconductor, semimetal and metal materials and predicting the 
band gap of the semiconductors. In our results, the dependence of the band gap 
on the $d$-orbital is explicit. In conclusion, we have investigated the properties 
of stable and metastable phases for a large set of TMD materials, and our findings 
may be auxiliary in the synthesis of metastable phases and in the development of 
new TMDs applications.	
\end{abstract}
	
\maketitle
	
\section{Introduction}\label{sec:Introduction}

Layered Materials have been known for \num{50} years \cite{Wilson_193_1969} and 
are applied in areas as diverse as dry lubricants \cite{Clauss_1972}, batteries 
\cite{Whittingham_4271_2004}, catalysts \cite{Lv_56_2014}, among others. 
Recently, 
layered transition-metal dichalcogenides (TMDs), materials with chemical 
formula i
\ce{$MQ$2}, with $M$ being a transition-metal and $Q$ a chalcogen (\ce{S}, 
\ce{Se} 
and \ce{Te}), have attracted wide technological interest due to their capacity 
of being isolated into one layer, like graphene does 
\cite{Chhowalla_263_2013,Geim_183_2007}, 
and the wide spectrum of the electronic properties they present, being metals, 
semimetals, semiconductors and insulators \cite{Kolobov_2016}. Recently, there 
were reported TMDs presenting exotic electronic properties, such as having 
topological 
insulator states \cite{Peng_65_2017}, being a Weyl semimetal 
\cite{Deng_1105_2016} 
and displaying charge density waves \cite{Kolobov_2016}. The wide spectrum of 
properties 
presented by TMDs is enabled by their large number of chemical compositions 
combining 
$M$ and $Q$ and the existence of several polymorphic phases 
\cite{Hulliger_1977,Chhowalla_263_2013,Kolobov_2016}.

In TMDs, layers composed of covalently bound $M$ and $Q$ planes are bound to 
each 
other by van der Waals interactions, and distinct coordination environments of 
the metal atoms within each layer generate structural polymorphism in these 
materials. 
Among these polymorphic phases, we highlight the most stable ones for a wide 
variety 
of materials \cite{Voiry_2702_2015,Hulliger_1977}: $(i)$ trigonal prismatic 
(2H), 
$(ii)$ octahedral (1T) and $(iii)$ distorted octahedral (1T$_{\text{d}}$). The 
lowest energy polymorph for a TMD depends mainly on the atomic radii and on the 
filling of the metal $d$- orbitals \cite{Voiry_2702_2015}, e.g., \ce{Ti} group 
metals (\ce{Ti}, \ce{Zr} and \ce{Hf}) favor 1T as the lowest energy phase 
\cite{Yang_931_2017}. 
On the other hand, the Peierls distortion mechanism is crucial for the 
energetic 
favoring of the 1T$_{\text{d}}$ phase in some compounds, since it breaks the 
degeneracy 
of electronic states, reducing the energy \cite{Yang_931_2017,Besse_2018}. 

The coexistence among different phases is linked to parameters such as 
temperature 
or pressure\cite{Keum_482_2015,Cho_625_2015}, and for the same material two 
different 
polymorphic phases may drastically change properties, e.g., \ce{MoS2} in 2H and 
1T phase is a semiconductor and a metal respectively 
\cite{Wypych_1386_1992,Tang_3743_2015}. 
The synthesis of many polymorphic phases has been possible with the advance of 
experimental techniques 
\cite{Wong_9_2016,Diaz_191606_2016,Loh_18116_2015,Tan_10584_2014}, 
allowing to obtain polymorphic phases which are not at the lowest energies. 
However, 
few studies were done in these metastable phases and a comprehensive 
characterization 
of TMD compounds and polymorphs is important to explore their properties and 
identify 
stability factors. 

A key factor for the renewed interest in layered TMDs is the production of 
two-dimensional 
(2D) materials from the mechanical or chemical exfoliation of the layers 
\cite{Novoselov_10451_2005}. 
The easiness to exfoliate the materials comes from the weak binding between 
layers, 
which depends on the van der Waals interactions, much weaker than the in-plane 
covalent bonding. However, some materials are more difficult to exfoliate than 
others due to stronger interlayer binding and, as suggested by Monet \textit{et 
al.} 
\cite{Mounet_246_2018}, the exfoliation energy can be used to determinate how 
easy 
it is to exfoliate the layers from the crystal. Previous studies report 
exfoliation 
energies only for TMDs at the lowest energy phases and studies involving 
another 
polymorphic phases are not common. Therefore, a thorough evaluation of the 
exfoliation 
energy in different TMD compounds and polymorphs is called for as an effective 
way to guide the production of two-dimensional materials.

To obtain a comprehensive description of the properties of layered TMDs, we 
performed 
a first-principles investigation of the stability, exfoliation energy and 
electronic 
properties of TMDs formed by \ce{Ti}-, \ce{V}- and \ce{Cr}-group transition 
metals 
and \ce{S}, \ce{Se} and \ce{Te}, in three different polymorphic phases: 2H, 1T, 
1T'. The elastic constants of the materials were calculated, which provide 
means 
to evaluate their stability, and the magnetic order was also considered. Based 
on the exfoliation energies and on the analysis of charge transfer between 
metals 
and chalcogens, we identified trends correlating the intralayer charge transfer 
with the magnitude of interlayer binding. Lastly, we classified all the studied 
TMDs compositions and polymorphic phases according to their electronic 
properties.

\section{Theoretical Approach and Computational Details}\label{sec:TheoreticalApproach}

Our first-principles calculations are based on the density functional theory 
(DFT) 
formalism \cite{Hohenberg_B864_1964,Kohn_A1133_1965} within the semi-local 
exchange 
and correlation functional proposed by Perdew--Burke--Ernzerhof (PBE) 
\cite{Perdew_3865_1996}. 
The Kohn--Sham equations were solved using the PAW method 
\cite{Blochl_17953_1994} 
as implemented in the Vienna \textit{ab initio} simulation package (VASP), 
version 
$5.4.1$ \cite{Kresse_13115_1993,Kresse_11169_1996}. We focus on layered 
dichalcogenides, 
and it is well known from the literature that the PBE functional underestimates 
long range interactions such as the London dispersion \cite{Tsuneda_2014}. To 
minimize 
this problem, we employed the semi-empirical DFT-D3 method proposed by Grimme 
\textit{et al.} 
\cite{Grimme_154104_2010}, which has been shown to provide structural 
properties 
for \ce{MoS2} in good agreement with experimental results 
\cite{Peelaers_305502_2014}.

It has been well known that (semi-) local functionals fail to accurately 
predict 
band gap energies 
\cite{Perdew_5048_1981,Parr_1994,MoriSanchez_146401_2008,Kim_035203_2009,Engel_2011,Bastos_65702_2018},
and hence, to minimize this problem, the electronic properties, such as density 
of states (DOS) and band structures were computed using the hybrid functional 
proposed 
by Heyd--Scuseria--Ernzerhof \cite{Heyd_8207_2003,Heyd_219906_2006} (HSE06), 
which 
contains the PBE correlation and separates the exchange term in long and short 
range terms by a screening function with the parameter $\omega = 
\SI{0.206}{\angstrom}^{-1}$. 
The short range term is composed of \SI{25}{\percent} of exact exchange and 
\SI{75}{\percent} 
of PBE exchange, while the long-range term is composed only by PBE exchange. We 
included also the relativistic effects of spin-orbit-coupling (SOC) for the 
valence states through the second-variational approach 
\cite{Koelling_3107_1977}. 

Spin-orbit coupling (SOC) effects were included for the valence states through
the second-variational approach\cite{Koelling_3107_1977}. As showed in our previous
work \cite{Bastos_65702_2018}, the SOC have small impact in the
structural properties, hence, for volume equilibrium, relative energy stability,
exfoliation energy and elastic constants calculations, the SOC was neglected.
For electronic properties, such as band structures and DOS, 
SOC corrections were considered only in combination with the PBE functional, 
due to the high computational cost of the combined HSE06-SOC calculations.

Structural optimizations were performed with PBE+D3 through the minimization of 
the stress tensor and of the forces on every atom. We used, for the plane-waves 
basis set, cutoff energy of $2\times$ the maximum energy recommended by VASP 
(ENMAX parameter from POTCAR file, as described in Table I of Supplemental 
Material) 
to determine the equilibrium lattice parameters. Using the optimized 
structures, 
the elastic constants were computed using $(i)$ contributions from 
strain-stress 
relations for distortions in the lattice with rigid ions and $(ii)$ ionic 
relaxation 
contributions, determined from the inversion of ionic Hessian matrix 
\cite{Page_104104_2002,Wu_035105_2005}. 
To achieve the convergence condition for the elastic constant we increased the 
cutoff energy to $2.5\times$ENMAX. The cutoff energy employed to compute the 
electronic properties, i.e., DOS, band structure and Bader charge, as well as 
to obtain cohesive energy and exfoliation energy, was $1.125\times$ENMAX. 

For the integration in the first Brillouin zone, we employed a Monkhorst--Pack 
scheme \cite{Monkhorst_5188_1976} using a \textbf{k}-mesh of 
$11{\times}11{\times}2$ 
for 2H-\ce{MoS2}, and meshes with same \textbf{k}-point density for the 
remaining 
structures, to obtain the equilibrium structure parameters. On the other hand, 
the \textbf{k}-mesh was increased in all systems, e.g. to 
$22{\times}22{\times}5$ 
for 2H-\ce{MoS2}, to compute the electronic properties. Due to the limitations 
of parallel calculations of elastic constants in VASP, we employed a 
$\Gamma$-centered 
\textbf{k}-mesh with fixed grid of $16{\times}16{\times}4$, 
$12{\times}12{\times}6$ 
and $10{\times}18{\times}5$ for 2H, 1T, and 1T$_\text{d}$ structures, 
respectively, 
for all chemical compositions. More details about the computational approach 
are 
provided in the Supplemental Material. 

\section{Crystal structures: \ce{$MQ$2}}                                     
                                                                                
\begin{figure*}                                                                 
        \centering                                                              
        \includegraphics[width=0.9\linewidth]{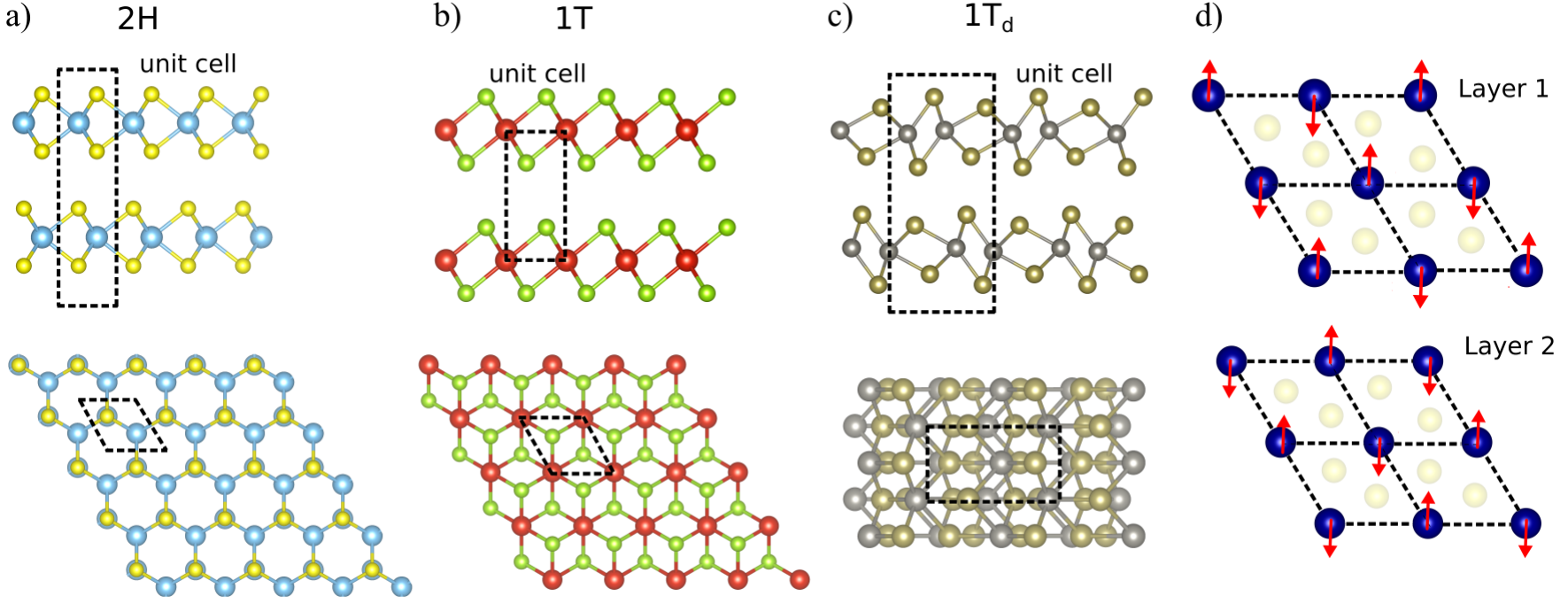}               
        \caption{Lateral and top view of the TMDs crystal structures of polytypes 
                a) Trigonal prismatic (2H), b) octahedral (1T) and c) distorted 
                octahedral (1T$_\text{d}$). The unit cells are represented in dashed 
                lines. d) Example of the supercell used to obtain the magnetic  
                ordering, where the red arrows represent the initial magnetic   
                moments in transition metal atoms.}                             
        \label{fig:CrystalStructures}                                           
\end{figure*}                                                                   
                                                                                
Our study concentrates on the most commonly observed TMD structural phases, 2H, 
1T and 1T$_\text{d}$\cite{Kolobov_2016}. In Fig. \ref{fig:CrystalStructures}, we  
present the schematics of the polytypes crystal structures, with \ref{fig:CrystalStructures}a), 
\ref{fig:CrystalStructures}b) and \ref{fig:CrystalStructures}c) showing the top 
and lateral views, and \ref{fig:CrystalStructures}d) indicating an example of supercell 
employed to access the magnetic ordering. The 2H structure, shown in \ref{fig:CrystalStructures}a), 
is composed of a hexagonal lattice, with 2 formula units (f.u.) per unit cell, 
whose atoms planes are in the AbA BaB stacking sequence (capital and lower case 
letters for chalcogen and metal atoms planes, respectively), belonging to the $P6_{3}/mmc$ 
space group \cite{Hulliger_1977}. The 1T structure, shown in \ref{fig:CrystalStructures}b), 
is composed of a hexagonal lattice with 1 f.u. per unit cell, with AbCABC stacking 
sequence, belonging to the $P\bar{3}m1$ space group \cite{Hulliger_1977}. Each 
layer of the 1T$_\text{d}$ structure, shown in \ref{fig:CrystalStructures}c), can 
be generated from a 1T monolayer by reconstructions in a $2\times1$ orthorhombic 
cell, originating dimerized lines of metal atoms, a distortion which has been shown 
to be driven by a Peierls transition mechanism \cite{Besse_2018}. The 1T$_\text{d}$ 
structure is composed of an orthorhombic lattice with 4 f.u. in the unit cell, 
belonging to the $Pnm2_{1}$ space group. The bonding geometry symmetries correspond 
to $D_{6h}$, $D_{3d}$ and $C_{2v}$ point groups for 2H, 1T and 1T$_\text{d}$, respectively.
                                                                                
We address materials composed by \ce{Ti}-, \ce{V}- and \ce{Mo}-group metals, with 
not fully occupied $d$-orbitals, thus some of the TMDs can exhibit non-zero magnetic 
moment, as has been experimentally observed in \ce{VS2} and \ce{VSe2} with ferromagnetic 
ordering in low temperatures \cite{Gao_5909_2013,Hulliger_1977}. To address the 
intrinsic magnetism in bulk TMDs, we employed supercells containing eight f.u., 
allowing to model antiferromagnetic configurations, as exemplified in Fig. \ref{fig:CrystalStructures}d). 
For non-magnetic and ferromagnetic orderings, the unit cell was employed, since 
it can represent such configurations. To increase the reliability of our results, 
we built four antiferromagnetic initial configurations with supercells, plus the 
ferromagnetic one, and equilibrium volumes were obtained for every initial configuration, 
from which the lowest energy structure was subsequently selected. Tables with the 
energy comparison for the five initial configurations can be found in the Supplemental 
Material.
                                                                                
\section{Results} \label{sec:Results}

\subsection{Relative energy stability}\label{subsection:RelativeStability}

\begin{figure*}
	\centering
	\includegraphics[width=0.9\linewidth]{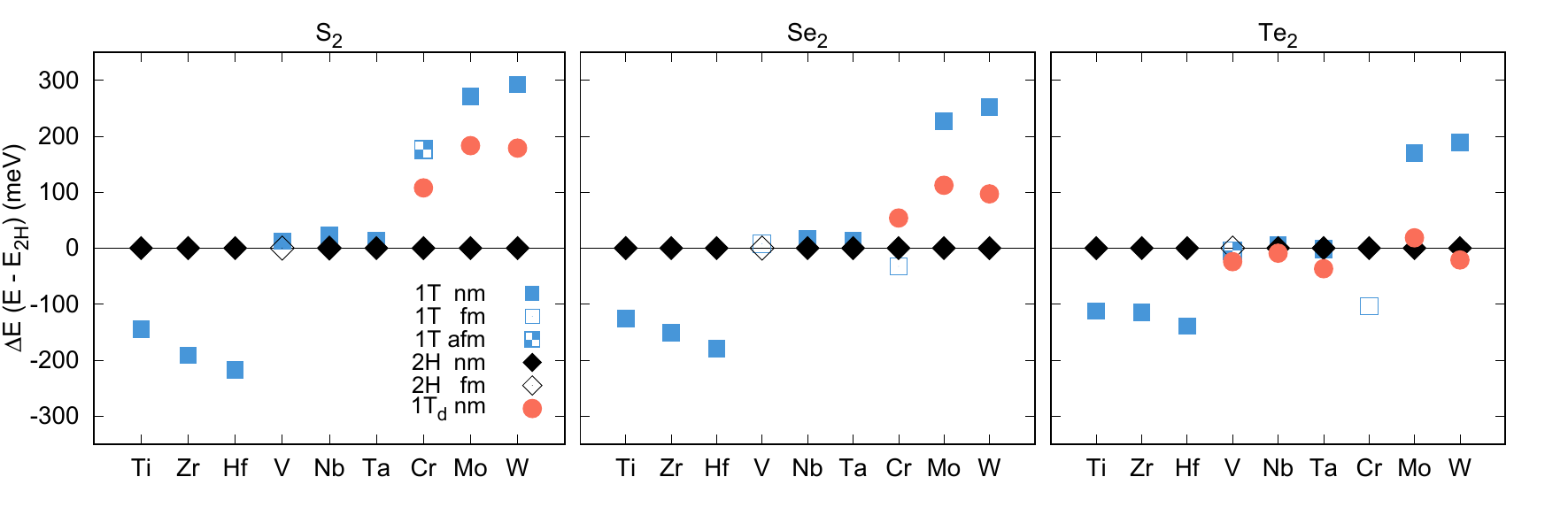}
	\caption{Relative total energy between 1T, 1T$_{\text{d}}$ and 2H phases 
		for \ce{$M$S2}, \ce{$M$Se2} and \ce{$M$Te2}, where $M$ is the metal 
		indicated in the x-axis. 2H phase is used as reference, so the vertical 
		axis presents the values for 
		($E_{(1\text{T},1\text{T}_{\text{d}},2\text{H})}-E_{2\text{H}}$). 
		The magnetic ordering is indicated by the symbols in the key: 
		nonmagnetic 
		(nm), ferromagnetic (fm) and antiferromagnetic (afm).}
	\label{fig:RelativeStability}
\end{figure*}

We obtained the equilibrium geometric configurations for all compounds and 
analyzed 
the relative stability between the phases (2H, 1T, 1T$_{\text{d}}$) by 
comparison 
of the total energy, Fig. \ref{fig:RelativeStability}. For \ce{Ti} group 
compounds, 
\ce{V} group selenides and tellurides, and \ce{CrTe2}, the 1T$_{\text{d}}$ 
phase 
does not present a local minimum structure in the potential surface, i.e., even 
if the structural relaxation starts from the 1T$_{\text{d}}$ structure, it 
yields 
the structural configuration of the 1T phase. Therefore, these compounds are 
not 
stable in the 1T$_{\text{d}}$ phase and these structures were not further 
considered 
in our calculations.

Figure \ref{fig:RelativeStability} shows the relative total energies obtained 
for 
the three polymorphic phases using 2H phase as the reference. For all the 
\ce{Ti} group 
compounds (\ce{Ti}, \ce{Zr} and \ce{Hf} combined with \ce{S},\ce{Se} and 
\ce{Te}) 
the lowest energy phase is 1T, as expected \cite{Chhowalla_263_2013}, while for 
the \ce{V} and \ce{Cr} groups compounds there is an alternation of the lowest 
energy 
phase. The \ce{V} group compounds have a small energy difference between the 
phases, 
which is manifested with the synthesis of 2H and 1T phases among these 
materials 
\cite{Kolobov_2016}, and 2H is the most stable for compounds with \ce{S} and 
\ce{Se} 
whereas 1T$_\text{d}$ is the most stable for compounds with \ce{Te}. Compounds 
with \ce{V} are experimentally observed in the 2H phase, in contrast with our 
results, 
however it has been shown that due to the small energy difference between the 
phases, 
temperature effects might change the lowest energy phase, in agreement with the 
synthesis of 1T-\ce{VS2} at room temperature\cite{Zhang_10821_2013}. 

In \ce{Cr} group, 2H predominates as the lowest energy phase (\ce{Mo$Q$2}, 
\ce{WS2}, 
\ce{WSe2}), as expected \cite{Chhowalla_263_2013}. The exceptions are \ce{CrSe2}
(1T), \ce{CrTe2} (1T), and \ce{WTe2} (1T$_\text{d}$), which were all 
experimentally 
observed crystal structures 
\cite{Freitas_014420_2013,Freitas_176002_2015,Kolobov_2016}. 
Finally, the room temperature crystal structures of \ce{NbTe2} and \ce{TaTe2}, 
which are formed of a monoclinic lattice \cite{Brown_264_1966}, are not 
considered
in our calculations. However, among the structures considered, the 
1T$_\text{d}$ 
phase, that has the same intra-layer structural configuration of distorted 
octahedral 
coordination of metal atoms, was obtained as the lowest energy one for these 
compounds.

Therefore, our results are in line with the general rule of the strong 
influence 
of the filling of metal $d$ orbitals on the lowest energy phase of each 
compound 
\cite{Chhowalla_263_2013,Yang_931_2017}, as can be seen in the preference of 1T 
phase for \ce{Ti} group compounds, and mostly 2H phase for TMDs of \ce{V} and 
\ce{Cr} 
groups. However, other effects are important to determine the energetic favored 
phase, like magnetism. From our results, only the \ce{V} group and \ce{Cr} 
compounds 
favor a magnetic ordering, depending on the phase: \ce{VS2}, \ce{VSe2}, 
\ce{VTe2} 
and \ce{NbS2} in 2H phase are ferromagnetic, as well as \ce{VSe2} and 
\ce{TaTe2} 
in 1T phase, while \ce{CrS2} and \ce{VTe2} in 1T phase are antiferromagnetic. 
This 
information is also presented in Fig. \ref{fig:RelativeStability}. A recent 
theoretical 
work \cite{Mounet_246_2018} reported that the 1T monolayers of \ce{V} 
dichalcogenides 
are ferromagnetic, and 1T-\ce{CrSe2} monolayer is reported as 
antiferromagnetic. 

The antiferromagnetic ordering was also obtained by experimental measurements 
in bulk 1T-\ce{CrSe2}\cite{Freitas_014420_2013}, although we obtained lower 
energy for the ferromagnetic ordering. The energy difference between the two 
orderings in our calculations, however, is of only \SI{4}{\milli\electronvolt}, 
i.e., the phases are approximately degenerate and stable. Furthermore, this 
difference is so small that, the use of a different vdW correction may change 
the result, also considering that variations on the $c$ lattice parameter were 
shown to modify the energetic preference between the two 
orderings\cite{Freitas_014420_2013}. The relative energies 
for all magnetic configurations are presented in the Supplemental Material.

 \subsection{Exfoliation energy}\label{subsection:MechanicalExfoliationEnergy}

\begin{figure*}
 	\centering
 	\includegraphics[width=0.9\linewidth]{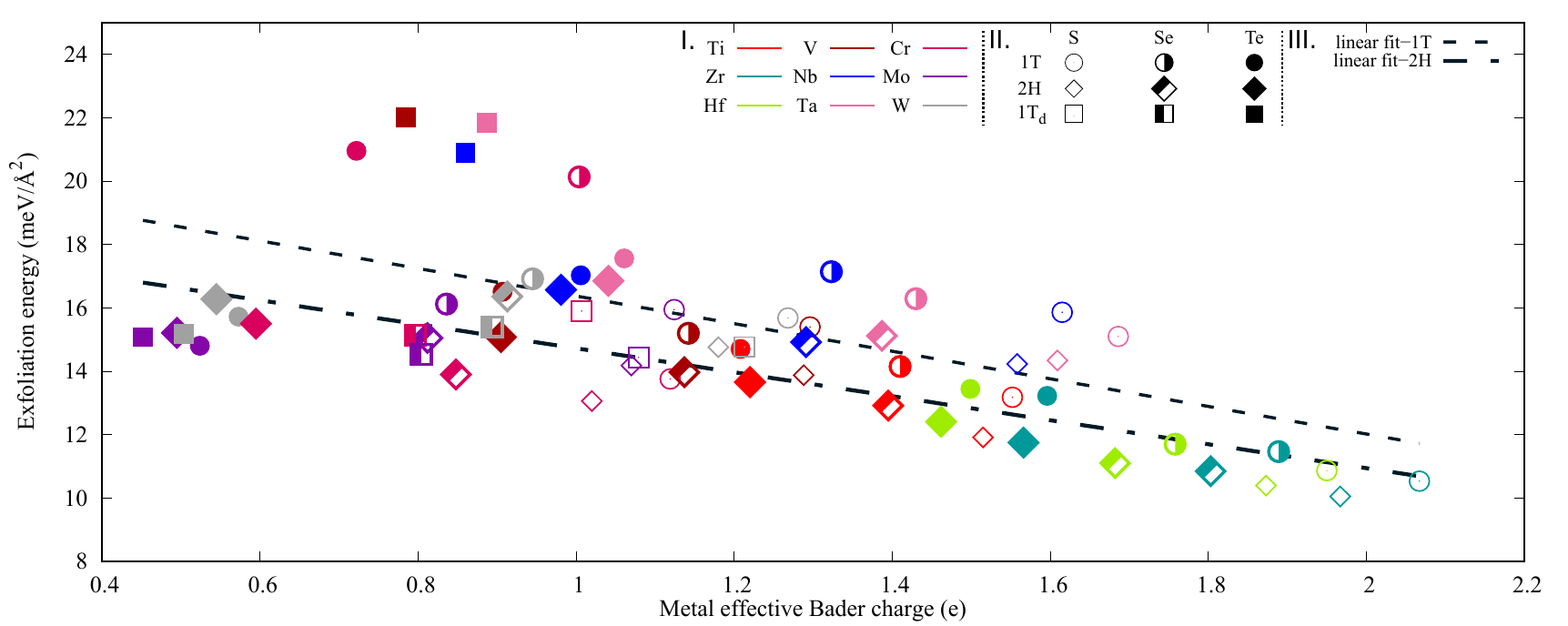}
 	\caption{Exfoliation energy as function of the metal effective Bader 
	charge. Compounds can be identified by the combination of the color in table I  with the symbol in table II. Table III presents a legend for the linear fitting of phases 1T and 2H.}
 	\label{fig:InterlayerEnergy}
 \end{figure*}

To investigate the strength of interlayer binding in TMDs and determine how easily 
they can be exfoliated, we calculated the exfoliation energy. As the polymorphic 
structures have different unit cells, we calculate exfoliation energies per monolayer 
area in the unit cell. In Fig. \ref{fig:InterlayerEnergy}, the exfoliation energy 
is shown as a function of metal effective Bader charge \cite{Bader_1994,Tang_084204_2009},
i.e., an estimate of the charge transfer from metal to chalcogen atoms. Several 
works, using different levels of vdW corrections \cite{Bjorkman_235502_2012,Bjorkman_424218_2012,Choudhary_5179_2017,Ashton_106101_2017,Mounet_246_2018},
propose a classification of the materials that are easily or potentially exfoliable 
based on the exfoliation energy. We adopt the classification in which materials 
with exfoliation energy of \SIrange[range-units=single]{15}{20}{\milli\electronvolt\per\angstrom^2} 
are considered easily exfoliable, while materials with energies above these values 
up to \SI{130}{\milli\electronvolt\per\angstrom^2} are considered potentially exfoliable. 

In our results, almost all the studied compounds have exfoliation energy in the 
range of \SIrange[range-units=single]{10}{17}{\milli\electronvolt\per\angstrom^2} 
and can be classified as easily exfoliable, as shown in Fig. \ref{fig:InterlayerEnergy}. 
The exceptions are: $(i)$ \ce{CrSe2} and \ce{CrTe2} in the 1T phase, with \SIlist[list-units=single]{20;21}{\milli\electronvolt\per\angstrom^2}, 
respectively; $(ii)$ \ce{VTe2}, \ce{TaTe2} and \ce{NbTe2} in the 1T$_\text{d}$ 
phase, with \SIlist[list-units=single]{22;22;21}{\milli\electronvolt\per\angstrom^2}, 
respectively. 

An analysis of Fig. \ref{fig:InterlayerEnergy}, shows that the exfoliation energy 
decreases linearly with the increase of the metal effective Bader charge, presenting 
a clear trend. This relation arises because with the increase of the charge on 
metal atoms, and consequently increase of the magnitude of the charge on the chalcogen 
plane, the effective Coulomb repulsion among the layers also increases, resulting 
in larger interlayer distances and smaller exfoliation energy. There is a small 
difference between polymorphic phases, as indicated by the separate linear fittings 
in Fig. \ref{fig:InterlayerEnergy}, which may be related to the atom ordering in 
the chalcogen plane. As generally 1T phase has higher in-plane lattice parameters 
when compared with 2H, less charge is accumulated in the chalcogen plane in 1T 
than in 2H. As a result, the distance between planes is smaller and the exfoliation 
energy is larger. Nevertheless, this difference is small, as shown for 2H 
and 1T \ce{ZrS2} in Fig. \ref{fig:InterlayerEnergy}. When the charge transfer is 
lower, Coulomb repulsion plays a smaller role, leading the exfoliation energies 
to be more dependent on other effects, e.g. van der Waals interaction, causing 
larger deviations from the linear trend, as with the cases of stronger ($>\SI{20}{\milli\electronvolt\per\angstrom^2}$) 
interlayer binding. Thus, the linear correlation is not clear for 1T$_{\text{d}}$, 
which may be due to the non-uniformities in the chalcogen plane caused by the distortions 
typical of this phase. All values of exfoliation energy and Bader charge are available 
in Supplemental Material.

\subsection{Equilibrium volume}\label{subsec:EquilibriumVolume}

\begin{figure*}
	\centering
	\includegraphics[width=0.75\linewidth]{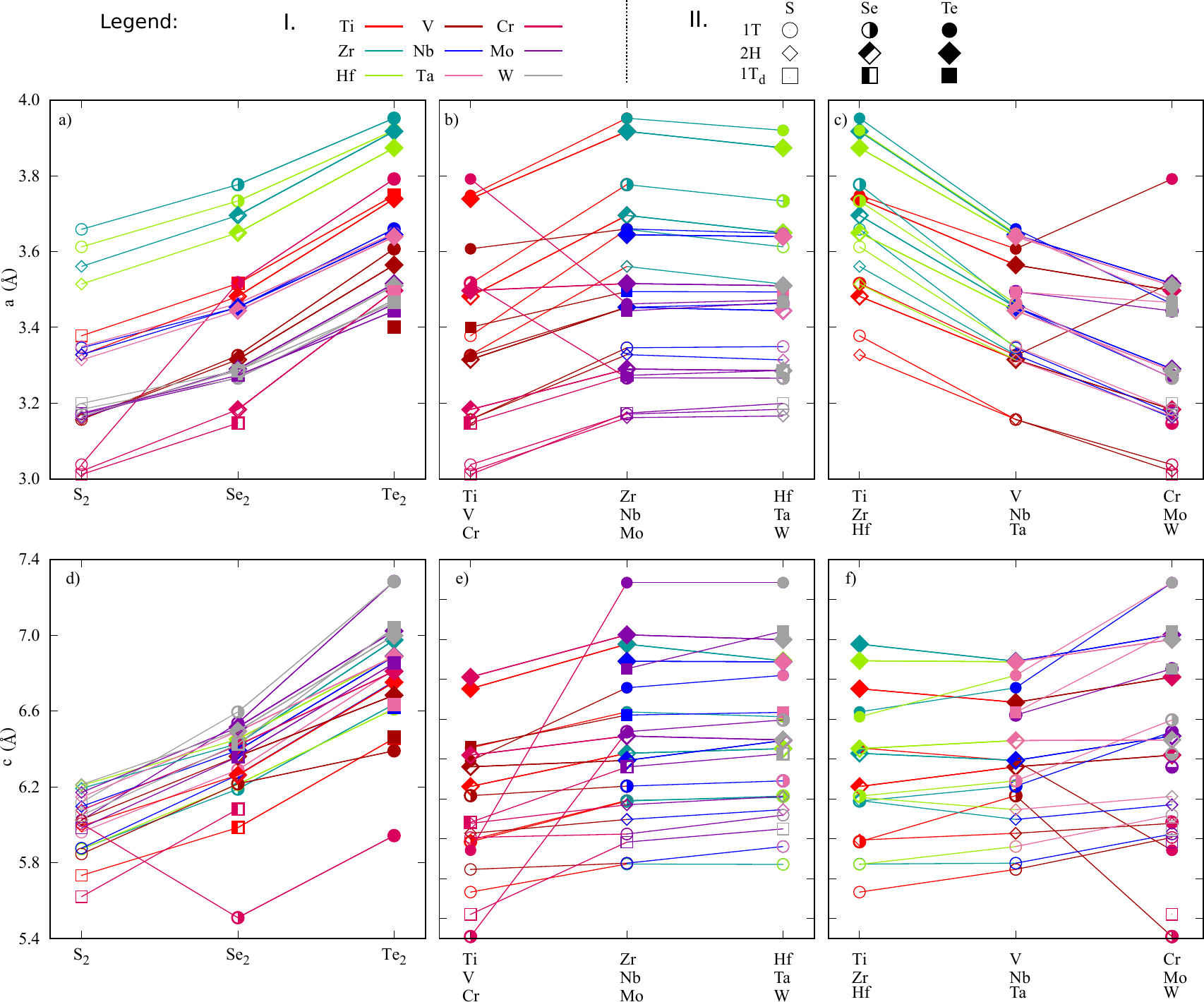}
	\caption{Lattice parameters of the 2H, 1T and 1T$_{\text{d}}$ TMDs. 
		Compounds are identified by the combination of the color, in table I, 
		with the symbols, in table II. The out-of-plane lattice parameter is 
		normalized by the number of layers in the unit cell. In a) and d) 
		the dependence on the chalcogen species is shown, b) and e) present 
		the dependence on the transition metal period, and c) and f) show 
		the dependence on the metal group. All numerical values are 
		presented in the Supplemental Material.}
	\label{fig:Lattice}
\end{figure*}

In Fig. \ref{fig:Lattice}, we show in-plane, $a$, and perpendicular, $c$, lattice 
parameters. Due to the different number of layers in the unit cells between the 
polymorphs, in order to compare all the materials, we used the value of the out-of-plane 
lattice parameter divided by the number of layers to obtain $c$. As expected, the 
lattice parameters increase monotonically with the chalcogen atomic radius (covalent 
radius reference values are \SIlist{1.04;1.14;1.32}{\angstrom} for \ce{S}, \ce{Se}, 
\ce{Te} respectively\cite{Kittel_2004}), as shown in Fig. \ref{fig:Lattice} a),d). 
The exceptions are the 1T \ce{Cr} compounds that present an abnormal increase of 
$c$ from \ce{CrSe2} to \ce{CrS2}.  

The effect of varying transition-metal period, with fixed metal group and 
chalcogen 
species, on the lattice parameters is represented in Fig. \ref{fig:Lattice} 
b), e). 
Compounds with transition-metal belonging to period \num{4} (\ce{Ti}, \ce{V} and 
\ce{Cr}) 
have the smallest parameters, while compounds with transition metals belonging 
to periods \num{5} (\ce{Zr}, \ce{Nb} and \ce{Mo}) and \num{6} (\ce{Hf}, \ce{Ta} and 
\ce{W}) 
have similar lattice parameters. This is also in agreement with the trend of 
atomic 
radii of the transition metals, which are the smallest for transition metals 
belonging 
to period \num{4} and have close values for transition metals from periods \num{5} and \num{6}, 
for 
example \SIlist{1.28;1.40;1.41}{\angstrom} for \ce{Cr}, \ce{Mo} and \ce{W}, 
respectively \cite{Kittel_2004}. 

\begin{table*}
	\centering
	\caption{Lattice parameters obtained from DFT, compared to their experimental values. Materials that are unstable in the phase 1T$_\text{d}$ are indicated. The experimental values were extracted from 	Ref.~\onlinecite{Hulliger_1977}. All values are presented in \si{\angstrom}.}
	\label{Table:LatticeParameters}
	\addtolength{\tabcolsep}{0pt}  
	\bgroup
	\def\arraystretch{1.2}
	\begin{tabular}{lllSSS@{\qquad}lllSSS@{\qquad}lllSSS}
		\toprule
		& phase            &         &{$a_0$} &{$b_0$}  & {$c_0$}   
		&             & phase            &          &{$a_0$}& {$b_0$} & 
		{$c_0$}    &              & phase             &           &{$a_0$}& 
		{$b_0$} & {$c_0$}    \\  \hline      
		\ce{TiS2}  & 2H              & DFT     & 3.33   &  3.33   & 11.99     
		&\ce{TiSe2}   & 2H              & DFT      & 3.48  & 3.48    & 
		12.51      &\ce{TiTe2}    & 2H               &  DFT      & 3.73  & 
		3.73    & 13.50       \\
		& 1T              &         & 3.38   &  3.38   &  5.73     
		&             & 1T              &          &  3.51 & 3.51    & 
		5.98       &              & 1T               &           & 3.74  & 
		3.74    & 6.45      \\
		& 1T$_{\text{d}}$ &         & \multicolumn{3}{c}{unstable} 
		&             & 1T$_{\text{d}}$ &          & 
		\multicolumn{3}{c}{unstable} &              & 1T$_{\text{d}}$  
		&           & \multicolumn{3}{c}{unstable} \\
		& 1T              & Exp.    & 3.41   & 3.41    & 5.70      
		&             & 1T              & Exp.     & 3.53  &  3.53   & 
		6.00       &              & 1T               & Exp.      & 3.76  &  
		3.76   & 6.52        \\ 
		\ce{ZrS2}  & 2H              & DFT     & 3.56   & 3.56    & 12.38     
		&\ce{ZrSe2}   & 2H              & DFT      & 3.69  & 3.69    & 
		12.85      &\ce{ZrTe2}    & 2H               & DFT       & 3.91  & 
		3.91    & 13.95       \\
		& 1T              &         & 3.65   & 3.65    & 5.87      
		&             & 1T              &          & 3.77  & 3.77    & 
		6.18       &              & 1T               &           & 3.95  & 
		3.95    & 6.63      \\
		& 1T$_{\text{d}}$ &         & \multicolumn{3}{c}{unstable} 
		&             & 1T$_{\text{d}}$ &          & 
		\multicolumn{3}{c}{unstable} &              & 1T$_{\text{d}}$  
		&           & \multicolumn{3}{c}{unstable} \\
		& 1T              & Exp.    & 3.66   &  3.66   & 5.81      
		&             & 1T              & Exp.     & 3.77  &  3.77   & 
		6.13       &              & 1T               & Exp.      & 3.95  &  
		3.95   & 6.63      \\  
		\ce{HfS2}  & 2H              & DFT     & 3.51   &  3.51   & 12.41     
		&\ce{HfSe2}   & 2H              & DFT      & 3.64  &  3.64   & 
		12.90      &\ce{HfTe2}    & 2H               & DFT       & 3.87  &  
		3.87   & 13.78       \\
		& 1T              &         & 3.61   &  3.61   &  5.87     
		&             & 1T              &          & 3.73  &  3.73   &  
		6.21      &              & 1T               &           & 3.91  &  
		3.91   &  6.61       \\
		& 1T$_{\text{d}}$ &         & \multicolumn{3}{c}{unstable} 
		&             & 1T$_{\text{d}}$ &          & 
		\multicolumn{3}{c}{unstable} &              & 1T$_{\text{d}}$  
		&           & \multicolumn{3}{c}{unstable} \\
		& 2H              & Exp.    & 3.37   &  3.37   & 11.78     
		&             & 2H              & Exp.     & 3.44  &  3.44   & 
		12.38      &              & 2H               & Exp.      &       
		&         &            \\  
		& 1T              &         & 3.95   &  3.95   & 6.65      
		&             & 1T              &          & 3.74  &  3.74   & 
		6.14       &              & 1T               &           & 3.95  &  
		3.95   & 6.65       \\
		\ce{VS2}   & 2H              & DFT     & 3.15   &  3.15   & 12.05     
		&\ce{VSe2}    & 2H              & DFT      & 3.31  &  3.31   & 
		12.72      &\ce{VTe2}     & 2H               & DFT       & 3.56  &  
		3.56   & 13.36      \\
		& 1T              &         & 3.15   &  3.15   &  5.84     
		&             & 1T              &          & 3.32  &  3.32   &  
		6.21      &              & 1T               &           & 3.60  &  
		3.60   &  6.39      \\
		& 1T$_{\text{d}}$ &         & \multicolumn{3}{c}{unstable} 
		&             & 1T$_{\text{d}}$ &          & 
		\multicolumn{3}{c}{unstable} &              & 1T$_{\text{d}}$  
		&           & 3.40  &  6.42   &  6.46     \\
		& 1T              & Exp.    &        &         &           
		&             & 1T              & Exp.     & 3.34  &  3.34   & 
		6.12       &              & 1T               & Exp.      &       
		&         &            \\  
		\ce{NbS2}  & 2H              & DFT     & 3.33   &  3.33   & 12.18     
		&\ce{NbSe2}   & 2H              & DFT      & 3.45  &  3.45   & 
		12.78      &\ce{NbTe2}    & 2H               & DFT       & 3.64  &  
		3.64   & 13.76       \\
		& 1T              &         & 3.34   &  3.34   &  5.87     
		&             & 1T              &          & 3.45  &  3.45   &  
		6.26      &              & 1T               &           & 3.65  &  
		3.65   &  6.76     \\
		& 1T$_{\text{d}}$ &         & \multicolumn{3}{c}{unstable} 
		&             & 1T$_{\text{d}}$ &          & 
		\multicolumn{3}{c}{unstable} &              & 1T$_{\text{d}}$  
		&           & 3.49  & 6.75    &  6.62     \\
		& 2H              & Exp.    &  3.31  &  3.31   & 11.88     
		&             & 2H              & Exp.     & 3.44  &  3.44   & 
		12.55      &              & 2H               & Exp.      &       
		&         &            \\  
		& 1T              &         &        &         &           
		&             & 1T              &          & 3.53  &  3.53   & 
		6.29       &              & 1T               &           &       
		&         &            \\
		\ce{TaS2}  & 2H              & DFT     &  3.32  &  3.32   & 12.16     
		&\ce{TaSe2}   & 2H              & DFT      & 3.44  &  3.44   & 
		12.98      &\ce{TaTe2}    & 2H               & DFT       & 3.32  &  
		3.32   & 11.98      \\
		& 1T              &         &  3.34  &  3.34   &  5.96     
		&             & 1T              &          & 3.46  &  3.46   &  
		6.29      &              & 1T               &           & 3.64  &  
		3.64   &  6.81      \\
		& 1T$_{\text{d}}$ &         & \multicolumn{3}{c}{unstable} 
		&             & 1T$_{\text{d}}$ &          & 
		\multicolumn{3}{c}{unstable} &              & 1T$_{\text{d}}$  
		&           & 3.49  &  6.68   &  6.63 \\
		& 2H              & Exp.    &  3.31  &  3.31   & 12.10     
		&             & 2H              & Exp.     & 3.43  &  3.43   & 
		12.72      &              & 2H               & Exp.      &       
		&         &       \\  
		& 1T              &         &  3.36  &  3.36   & 5.90      
		&             & 1T              &          & 3.47  &  3.47   &  
		6.27      &              & 1T               &           &       
		&         &       \\
		\ce{CrS2}  & 2H              & DFT     &  3.02  &  3.02   & 12.15     
		&\ce{CrSe2}   & 2H              & DFT      & 3.18  &  3.18   & 
		12.83      &\ce{CrTe2}    & 2H               & DFT       & 3.49  &  
		3.49   & 13.62 \\
		& 1T              &         &  3.31  &  3.31   &  5.34     
		&             & 1T              &          & 3.51  &  3.51   &  
		5.51      &              & 1T               &           & 3.79  &  
		3.79   &  5.94 \\
		& 1T$_{\text{d}}$ &         &  3.01  & 5.53    &  5.62     
		&             & 1T$_{\text{d}}$ &          & 3.14  &  5.75   &   
		6.08     &              & 1T$_{\text{d}}$  &           
		&\multicolumn{3}{c}{unstable} \\
		\ce{MoS2}  & 2H              & DFT     & 3.16   &  3.16   & 12.34     
		&\ce{MoSe2}   & 2H              & DFT      &  3.29 &  3.29   & 
		13.03      &\ce{MoTe2}    & 2H               & DFT       & 3.51  &  
		3.51   & 14.04 \\
		& 1T              &         & 3.17   &  3.17   &  6.02     
		&             & 1T              &          &  3.26 &  3.26   &  
		6.53      &              & 1T               &           & 3.46  &  
		3.46   &  7.28 \\
		& 1T$_{\text{d}}$ &         & 3.17   &  5.71   &  5.98     
		&             & 1T$_{\text{d}}$ &          &  3.27 &  5.94   &  
		6.35      &              & 1T$_{\text{d}}$  &           & 3.44  &  
		6.37   &  6.85 \\
		& 2H              & Exp.    & 3.15   &  3.15   & 12.29     
		&             & 2H              & Exp.     &  3.29 &  3.29   & 
		12.90      &              & 2H               & Exp.      & 3.51  &  
		3.51   & 13.97 \\  
		\ce{WS2}   & 2H              & DFT     & 3.16   &  3.16   & 12.42     
		&\ce{WSe2}    & 2H              & DFT      &  3.28 &  3.28   & 
		12.99      &\ce{WTe2}     & 2H               & DFT       & 3.50  &  
		3.50   & 13.99 \\
		& 1T              &         & 3.18   &  3.18   &  6.11     
		&             & 1T              &          &  3.26 &  3.26   &  
		6.59      &              & 1T               &           & 3.37  &  
		3.37   &  5.73 \\
		& 1T$_{\text{d}}$ &         & 3.19   & 5.70    &  6.05     
		&             & 1T$_{\text{d}}$ &          &  3.28 &  5.92   &  
		6.42      &              & 1T$_{\text{d}}$  &           & 3.46  &  
		6.27   &  7.04 \\
		& 2H              & Exp.    & 3.17   &  3.17   & 12.36     
		&             & 2H              & Exp.     &  3.28 &  3.28   & 
		12.95      &              & 1T$_{\text{d}}$  & Exp.      & 3.477 &  
		6.25   &  7.01 \\  
		\hline \hline
	\end{tabular}
	\egroup
	\addtolength{\tabcolsep}{1pt}  
	\label{tab:LatticeParameters}
\end{table*}

The atomic radii of transition metals (e.g., \SI{1.46}{\angstrom},
\SI{1.35}{\angstrom} and \SI{1.28}{\angstrom}  for \ce{Ti}, \ce{V} and \ce{Cr}, respectively \cite{Kittel_2004}) also
determine the decrease of the parameter $a$ with the increase of the column number of the transition metal group, for a fixed period and a chalcogen species,
as shown in Fig. \ref{fig:Lattice}c). There is no clearly defined trend for
the $c$ parameter. Compared with the experimental data for the 
already synthesized TMDs, as showed in the Table \ref{Table:LatticeParameters}, the calculated lattice parameters present mean absolute percentage errors (MAPE)
lower than \SI{1}{\percent} for in-plane lattice parameters ($a$ and $b$)
and lower than \SI{2}{\percent} for the perpendicular lattice parameter ($c$), indicating that PBE-D3 predicts reasonable values for VdW effects in TMDs.

\subsection{Elastic constants}\label{subsection:ElasticConstants}

As we consider some compositions and polymorphic phases not yet synthesized, 
the 
structural stability of the materials was addressed by the Born elastic 
stability 
criteria \cite{Born_1998}, and hence, we analyzed the stability of the TMDs 
through 
the evaluation of the elastic constants and verification of the Born elastic 
stability 
criteria as discussed by Mouhat and Coudert \cite{Mouhat_224104_2014}. Due to 
their 
symmetry, the crystals of 1T, 2H and 1T$_{\text{d}}$ phases present $6$, $5$ 
and 
$9$ non-zero and independent elastic constants, respectively, which must 
satisfy 
the necessary and sufficient conditions for stability discussed below. 

For the 1T phase ($P\bar{3}m1$ space group), with elastic constants $C_{11}$, 
$C_{12}$, 
$C_{13}$, $C_{14}$, $C_{33}$ and $C_{44}$ (and $C_{66} = (C_{11} - C_{12})/2$),
the conditions are: 
\begin{equation}
\begin{split}
C_{11} > |C_{12}|, \quad \quad & C_{44} > 0, \\
C^{2}_{13} < \frac{1}{2}C_{33}(C_{11} + C_{12}), \quad \quad & 
C_{14}^{2} < \frac{1}{2}C_{44}(C_{11} - C_{12})~.
\end{split}  
\end{equation}
For the crystal of 2H phase ($P6_{3}/mmc$ space group), that has the elastic 
constants 
$C_{11}$, $C_{12}$, $C_{13}$, $C_{33}$, $C_{44}$ (with $C_{66} = (C_{11} - 
C_{12})/2$), 
the conditions are:
\begin{equation}
C_{11} > |C_{12}|, \quad C^{2}_{13} < \frac{1}{2}C_{33}(C_{11} + C_{12}), \quad 
C_{44}>0. \\
\end{equation}
Finally, the elastic constants of the 1T$_\text{d}$ phase crystal ($Pnm2_{1}$ 
space 
group), $C_{11}$, $C_{12}$, $C_{13}$, $C_{22}$, $C_{23}$, $C_{33}$, $C_{44}$, 
$C_{55}$ 
and $C_{66}$, must satisfy the following conditions:
\begin{equation}
\begin{split}
\left[C_{11}C_{22}C_{33}+2C_{12}C_{13}C_{23}\right.&-C_{11}C_{23}^{2}\\ 
\ &\left.-C_{22}C_{13}^2-C_{33}C_{12}^2\right]>0,\\
C_{11}C_{12}>C_{12}^{2}, \quad\quad C_{11}>0,&\quad \quad \quad C_{44}>0,\\ 
C_{55}>0, & \quad \quad \quad C_{66}>0.
\end{split}
\end{equation}

\begin{figure}
	\centering
	\includegraphics[width=0.95\linewidth]{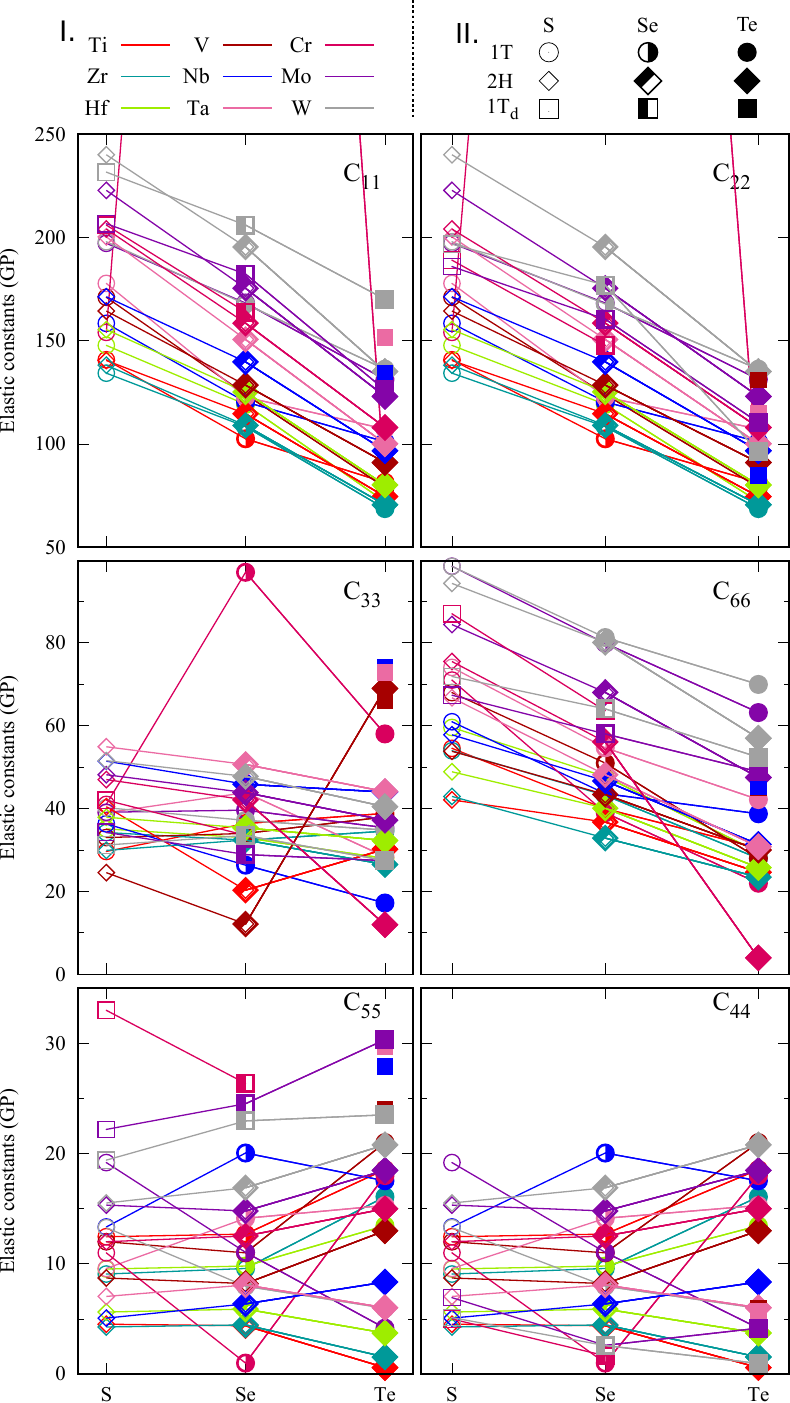}
	\caption{Diagonal elements of the elastic matrix of the 1T, 2H and 
		1T$_{\text{d}}$ TMD bulks.
	}
	\label{fig:Elastic}
\end{figure}

All the elastics constants values are shown in the Supplemental Material, while 
the diagonal elastic constants are shown in Fig. \ref{fig:Elastic}. We found 
that all conditions for the elastic stability are satisfied, and hence, those 
configurations are local minimum structures. Our results are in agreement with 
previous calculations 
\cite{Peelaers_12073_2014} and experimental results 
\cite{Kolobov_2016,Feldman_1141_1976}. 
For the elastic constants $C_{11}$ and $C_{22}$, which are not related to the 
out 
of plane direction ($z$), the magnitude decreases with the chalcogen radius and 
have higher values when compared with the other elastic constants, as shown in 
Fig. \ref{fig:Elastic}. This occurs because in plane binding, that is dominated 
by covalent bonds, is weaker for larger chalcogen radius, and is stronger than  
the out of plane van der Waals interactions. Therefore, the other elastic 
constants,
which are related to the $z$ direction, have a smaller magnitude, and present 
deviant
trends for the chalcogen radius, due to the role of the van der Waals 
interactions
in the interlayer interactions.

\subsection{Band structure and density of 
states}\label{subsection:BandStructure}

\begin{figure}
	\centering
	\includegraphics[width=0.9\linewidth]{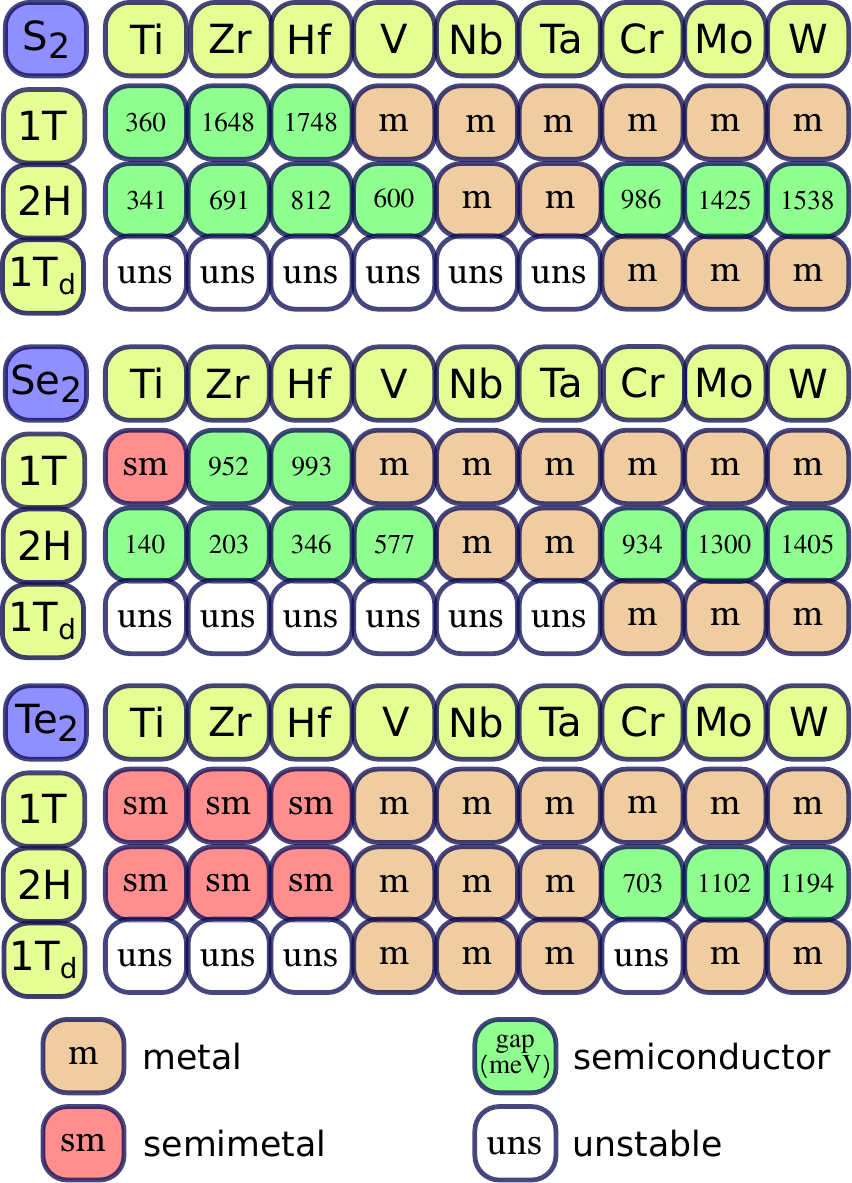}
	\caption{Classification of stable material phases according to their 
		electronic band gap: metal (m), semimetal (sm) or semiconductor. In 
		semiconductors, predicted gap energy values are shown in 
		\si{\milli\electronvolt} units.    Unstable phases are indicated by
		(uns).}
	\label{fig:ElectronicClassify}
\end{figure}

To characterize the materials according to their electronic properties, we 
calculated the band structure and the density of states of the \num{64} stable 
TMDs with the hybrid functional HSE06. The results for 2H-phase selenides of 
the $3d$-metals 
are shown in Fig. \ref{fig:BandStructure}, and the results for the other 
systems are in the Supplemental Material. From the analysis of the results, the 
materials were classified as metals, semi-metals or semiconductors, as 
indicated in Fig. 
\ref{fig:ElectronicClassify}. Among all the studied TMDs, \num{22} were 
identified 
as semiconductors, with band gaps ranging from 
\SIrange{0.20}{1.75}{\electronvolt}.

\begin{figure*}
	\centering
	\includegraphics[width=0.9\linewidth]{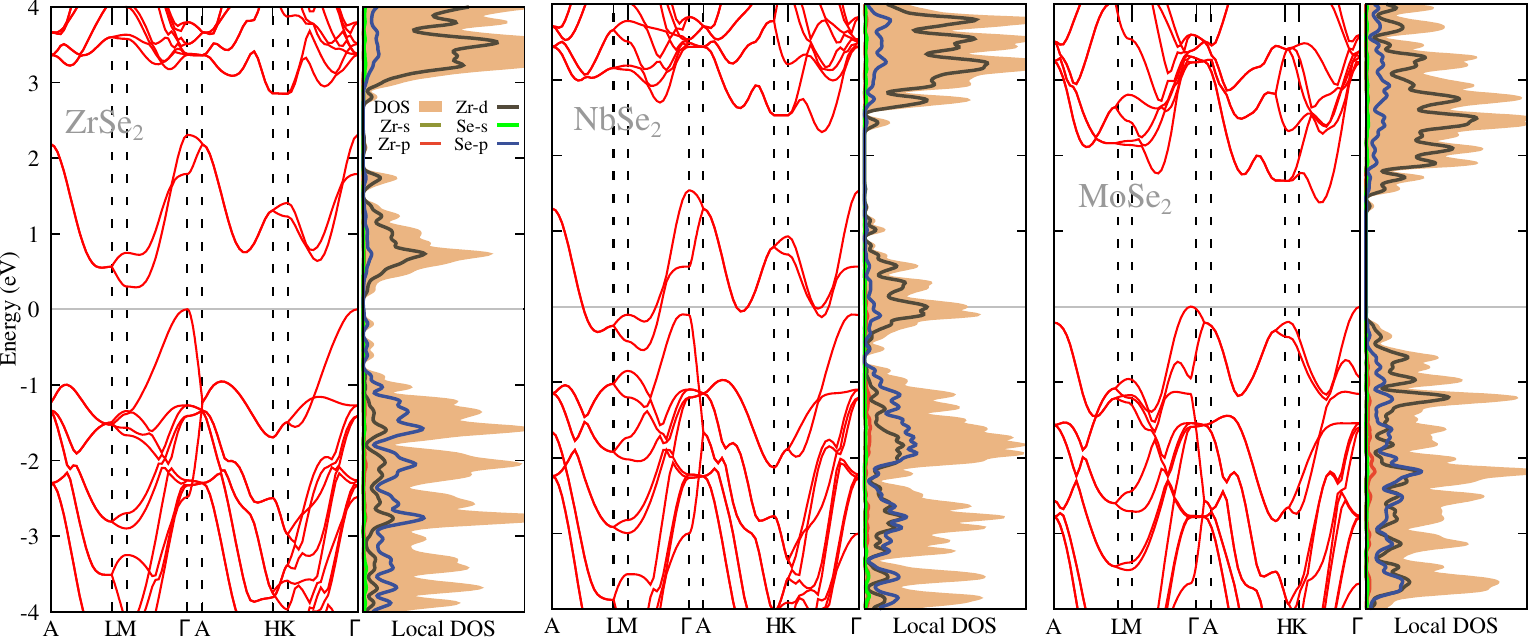}    
	\caption{2H-phase DFT-HSE06 band structures and densities of 
		states of \ce{ZrSe2} (left), \ce{NbSe2} (center) and \ce{MoSe2} 
		(right). Dashed lines indicate high-symmetry points in the first 
		Brillouin zone. At the right panels, shadowed curves indicate the 
		Density of States (DOS) while the solid lines indicate the local 
		density of states in $s$-, $p$- and $d$-orbitals of each atomic 
		species.}
	\label{fig:BandStructure}
\end{figure*}


In the literature, works that estimate the band gaps for TMDs
in the bulk phase mainly use the crystal structures acquired from 
crystallographic databases \cite{Lebegue_031002_2013,Zhu_081001_2018} 
presenting band gap values that	differ from our 
results. The difference of values is due to the use of the PBE 
functional, that is known to underestimate the band gap 
\cite{Perdew_5048_1981,Bastos_105002_2016,Bastos_65702_2018}. 
In Table \ref{table:gapsvalues}, we present the values of PBE, 
PBE+SOC,  HSE06 and experimental band gaps for the  
TMD semiconductor materials. 
The comparison of PBE and PBE+SOC shows that the inclusion of SOC 
modify the gap values usually from less than one to a few decades of 
\si{\milli\electronvolt}, exception made for \ce{ZrSe2}-1T, \ce{HfSe2}-1T 
and \ce{WTe2}-2H where this difference is about 150 \si{\milli\electronvolt}. 
However, the PBE functional usually underestimates 
the gap in the order of several hundreds of \si{\milli\electronvolt}s. 
Despite the absence of the SOC corrections on the HSE06 calculations this 
functional shows more realistic band gaps when compared with the experimentally 
measured values, preventing the high computational costs associated to the 
combined use of HSE06 and SOC.

For each phase and considering compositions with the same chalcogen, the 
transition metal $d$ electron count determines the electronic character of the 
material. For example, in 2H-\ce{$Q$Se} compounds, as shown in Fig. 
\ref{fig:BandStructure}, 
with the progressive filling of the $d$ band from \ce{Zr} to \ce{Nb} to \ce{Mo} 
\ce{ZrSe2} is a semiconductor, while \ce{NbSe2} is a metal, and the band is 
fully 
occupied in \ce{MoSe2}, recovering the semiconductor character. Therefore, if 
the 
$d$-orbitals are completely occupied or empty, the TMDs have semiconducting 
behavior, 
while if the $d$-orbitals have the partial occupation, the TMDs have conducting 
behavior. 
These results are in agreement with other reports in the literature 
\cite{Yang_931_2017,Voiry_2702_2015}. 
Because the crystal symmetry, i.e. the polymorphic phase, strongly affects the 
energy of the $d$ bands, the same compound can have different electronic 
properties depending on the polymorphic phase. For example, \ce{MoS2} is 
metallic in the 1T 
phase, but is a semiconductor in the 2H 
phase\cite{Wypych_1386_1992,Tang_3743_2015}.

Band gaps vary with the composition in a similar way for the three studied 
phases. 
The increase in chalcogen atomic number narrows the band gap, because the 
energy of $Q$-$p$ derived states, which compose the valence band maximum, is 
increased. 
For semiconductors with the same phase and chalcogen, the band gap increases 
with the 
transition metal atomic number, e.g., 
\ce{TiQ2}$^{gap}<$\ce{ZrQ2}$^{gap}<$\ce{HfQ2}$^{gap}$ 
in phase 1T, as shown in Fig. \ref{fig:ElectronicClassify}. This trend occurs 
due 
to the localization of the $d$ orbitals, since their energy difference from the 
Fermi energy increases, i.e., looking for compounds with the same phase, the 
$d$-orbitals 
in \ce{TiS2} have energies closer to Fermi energy than those in \ce{ZrS2}. 

\begin{threeparttable}
	\centering
	\caption{PBE, PBE with spin-orbit coupling ({PBE+SOC}), HSE06 and experimental band gaps for 
		TMD semiconductor materials. All energies are given in 
		\si{\milli\electronvolt}. }
	\label{}
	\begin{tabular}{lllSSSr}
		\hline
		\hline
		\toprule
		Material & Phase&& {PBE}  &  {PBE+SOC} &  {HSE06} & {Exp.} \\
		\hline
		\midrule
		\ce{TiS2} &1T   && 0      &  15        & 360   &    \\
		\ce{ZrS2} &1T   && 848    &  850       & 1648  &    \\
		\ce{ZrSe2}&1T   && 260    &  109       & 952   &    \\
		\ce{HfS2} &1T   && 988    &  974       & 1748  &    \\
		\ce{HfSe2}&1T   && 378    &  203       & 993   &    \\
		\ce{TiS2} &2H   &&  0     &  0         & 341   &    \\  
		\ce{TiSe2}&2H   &&  0     &  0         & 140   &    \\
		\ce{ZrS2} &2H   &&  0     &  0         & 691   &    \\
		\ce{ZrSe2}&2H   &&  0     &  0         & 203   &    \\
		\ce{HfS2} &2H   &&  159   &  172       & 812   &    \\
		\ce{HfSe2}&2H   &&  0     &  0         & 346   &    \\
		\ce{VS2}  &2H  &&  0     &  0         & 600   &    \\
		\ce{VSe2} &2H   &&  0     &  0         & 577   &    \\
		\ce{CrS2} &2H   &&  604   &  600       & 986   &    \\
		\ce{CrSe2}&2H   &&  600   &  589       & 934   &    \\
		\ce{CrTe2}&2H   &&  327   &  313       & 703   &    \\
		\ce{MoS2} &2H   &&  918   &  912       & 1425  &  1230\tnote{b}\\
		\ce{MoSe2}&2H   &&  870   &  860       & 1300  &  1090\tnote{b}\\
		\ce{MoTe2}&2H   &&  739   &  720       & 1102  &   880\tnote{a}\\
		\ce{WS2}  &2H  &&  1040  &  990       & 1538  &  1350\tnote{b}\\
		\ce{WSe2} &2H   &&  959   &  882       & 1405  &  1200\tnote{b}\\
		\ce{WTe2} &2H   &&  757   &  603       & 1194  &    \\
		\bottomrule
		\hline                                    
		\hline
	\end{tabular}
	\begin{tablenotes}
		\item[a] Scanning tunneling spectroscopy and ionic-liquid gated transistors at room temperature [\onlinecite{Lezama_021002_2014}]
		\item[b] Photocurrent spectra at room temperature [\onlinecite{Kam_463_1982}] 
	\end{tablenotes} 
	\label{table:gapsvalues}
\end{threeparttable}

\section{Conclusion}\label{section:Conclusions}

We investigated \num{27} TMD bulk compounds obtained by the combination of nine 
transition-metals (\ce{Ti}, \ce{Zr}, \ce{Hf}, \ce{V}, \ce{Nb}, \ce{Ta}, 
\ce{Cr}, 
\ce{Mo} and \ce{W}) with three chalcogens (\ce{S}, \ce{Se} and \ce{Te}) in 
three 
polymorphic phases, namely, 1T, 2H, and 1T$_{\text{d}}$. We obtained the 
equilibrium 
geometry configuration and the lowest energy phase for each material, which are 
in good agreement with experimental data for the already synthesized 
compositions. 
The magnetic ordering was also addressed, and some of the materials with 
transition-metal 
from the \ce{V} group and compounds with \ce{Cr} showed ferromagnetic or 
antiferromagnetic behavior. The effects of chemical composition on the 
equilibrium lattice parameters mostly follow the expected trends based on the 
atomic radius. 

To investigate the stability of the crystal structures, we obtained the elastic 
constants and employed the Born elastic stability criteria, which was satisfied 
for all the systems. The exfoliation energy of all stable materials was 
calculated, 
indicating that the majority of the studied TMDs have weak interlayer binding 
and 
therefore are predicted as easy to exfoliate in order to obtain their 
two-dimensional 
form. We found that the increase of the charge transfer within each layer 
decreases the magnitude of the exfoliation energy, due to the Coulomb repulsion 
between chalcogen planes. The electronic band structure and density of states 
were calculated, which allowed classifying the materials like metal, semimetal 
or semiconductor, according to their band gap. We demonstrated that the 
occupation of metal $d$ band determines the electronic character of the 
material. This study provides a comprehensible understanding of the properties 
of layered TMDs in different polymorphic phases, 
including material not yet synthesized, and therefore can contribute to further 
development of layered and two-dimensional materials based on TMDs. 

\section{Acknowledgments}
The authors gratefully acknowledge support from FAPESP (S\~ao Paulo Research 
Foundation,  
Grant Number 2017/11631-2), Shell and the strategic importance of the support 
given  by ANP (Brazil's National Oil, Natural Gas and Biofuels Agency) through 
the R\&D  
levy regulation. R.B. acknowledges financial support (Ph.D. fellowship) from 
FAPESP, 
Grant No. 2017/09077-7. This study was financed in part by the 
Coordena\c{c}\~ao de
Aperfei\c{c}oamento de Pessoal de N\'ivel Superior - Brasil (CAPES) - Finance
Code 001. G.M.S. acknowledges CAPES-CsF (grant No. 88881.068174/2014-01) and 
CNPq (grants No. 304289/2015-9 and 308806/2018-2).

\newpage


%

\end{document}